\documentclass[12pt, titlepage, reqno]{article}
\usepackage[super,comma,sort&compress]{natbib}
 \usepackage{amssymb, latexsym}
 \usepackage{amsmath} 
 \usepackage{amsthm}
 \usepackage{bm}
 \usepackage[mathscr]{eucal}
 \usepackage{enumerate}

 \renewcommand{\section}[1]{\medskip \addtocounter{section}{1}\raggedright 
     \textbf{\Roman{section}. \ #1}\medskip \setcounter{subsection}{0}
    \setlength{\parindent}{5ex}
 }
 \renewcommand{\subsection}[1]{\medskip \addtocounter{subsection}{1}\raggedright
    \textbf{\Alph{subsection}. \ #1} \medskip \setcounter{subsubsection}{0}\setlength{\parindent}{5ex}
}
 \renewcommand{\subsubsection}[1]{\medskip \addtocounter{subsubsection}{1}\raggedright
    \textbf{\arabic{subsubsection}. \ #1} 
}

 \theoremstyle{plain}
\usepackage{caption}
\usepackage{textcomp}
\usepackage{epstopdf}
\usepackage{graphicx}

\newcommand{\br}{\bar{r}}

\def\Xint#1{\mathchoice
{\XXint\displaystyle\textstyle{#1}}%
{\XXint\textstyle\scriptstyle{#1}}%
{\XXint\scriptstyle\scriptscriptstyle{#1}}%
{\XXint\scriptscriptstyle\scriptscriptstyle{#1}}%
\!\int}
\def\XXint#1#2#3{{\setbox0=\hbox{$#1{#2#3}{\int}$}
\vcenter{\hbox{$#2#3$}}\kern-.5\wd0}}

\def\dashint{\Xint-}

\newcommand{\vv}[1]{\bar{#1}}

\begin{document}

 \begin{titlepage}

 \begin{center}

 \textbf{A Quasianalytical Time Domain Mie Solution for Scattering from a Homogeneous Sphere}

 \vspace{10ex}

Jie Li\footnote{e-mail: jieli@egr.msu.edu}, Daniel Dault, and Balasubramaniam Shanker\\
 \vspace{2ex}
 Department of Electric and Computer Engineering \\
 
 Michigan State University \\

 East Lansing, Michigan 48823
 \end{center}

 \end{titlepage}



\begin{abstract}

A transient Mie-like solution for acoustic scattering from a spherical object is derived within a mesh-free and singularity-free Time Domain Integral Equation (TDIE) framework for the sound-soft, sound-rigid and penetrable cases.  The method is based on an expansion of the time domain Green's function that allows independent evaluation of spatial and temporal convolutions.  Solution of the TDIE system may be effected by descretizing the integral equations in space and time, forming a matrix system via the Method of Moments, and solving the system with the Marching on in Time algorithm.  Spatial discretization using tesseral harmonics leads to closed form expressions for spatial integrals, and use of a strictly band limited temporal interpolant permits efficient, accurate computation of temporal convolutions utilizing numerical quadrature.  The accuracy of these integrations ensures late time stability and accuracy of the deconvolution data.  Results presented demonstrate the accuracy and convergence of the approach for broadband simulations against Fourier transformed analytical data.  

\end{abstract}
 \addtocounter{page}{2}

\section{Introduction}

Analysis of wave scattering from spherical objects arises in many areas of acoustics, electromagnetics, biomedical imaging, and photonics research \citep{Henyey1991,Kozasa1992,Quinten1993,Dejong1993,Shankar1999}.  In many real-world applications, the quantities of interest are related to the transient scattering profile of the object, or to behavior that is not easily or efficiently characterized in frequency domain, such as broadband or nonlinear response \citep{Shi2000,Simpson1999}. Due to the various needs to model those transient phenomena directly in time domain, development of time domain modeling techniques is necessary. 

The analytical solution of scattering from spheres has been extensively studied for the time harmonic case, most commonly using multipole expansions such as those employed in Mie theory \citep{Perelman1978,Ishimaru1990} and transition matrix (T-matrix) theory \citep{Waterman1969,Gumerov2002,Ganesh2013}. Compared to the wide array of frequency domain methods, there is a relative dearth of research on analytical or semianalytical approaches for transient wave scattering from spherical objects. This disparity arises primarily from the difficulty of translating the traditional frequency domain methods directly into time domain, as direct inversion from the frequency domain of Mie-type solutions is not analytically tractable due to the singularity of Hankel functions. Time domain methods \citep{Ergin1998,Haggblad2012,HaDuong2003} have been studied extensively to handle targets with arbitrary geometry, these methods are mesh-based schemes often incurring impracticably large computational cost due to the large number of degrees of freedom induced by basis sets tied to meshes. They also introduce inaccuracy as a consequence of nonconformality of the mesh to the true underlying spherical geometry, and suffer from well-known stability problems. The contribution of the present work is a semianalytical method that is capable of capturing all relevant transient acoustic scattering phenomena without recourse to simplified models, while retaining the efficiency and accuracy afforded by frequency domain Mie-type expansions in the context of a Time Domain Integral Equation (TDIE) framework.  The resulting scheme is exact in space and exponentially accurate in time, of linear cost in both storage and computational complexity, and highly stable when solved using the Marching on in Time (MOT) method.


Expansion of transient multipoles has been examined in the literature \citep{Shlivinski1999,Heyman1996}, but most of the work has focused on the radiation problem; i.e., fields radiated due to a source distribution, and not on the scattering problem. To the authors' knowledge, the only previous work on the scattering problem is from Buyukdura and Koc \citep{Buyukdura1997} who sought a scattering solution for the sound-soft case by using a multipole expansion in terms of time-dependent spherical wave functions. In their work, time domain coefficients for those wave functions are obtained by matching the same mode on both sides of the system. However, an inverse Fourier transform of spherical Hankel function leads to unstable numerical implementation due to the presence of convolution involving Ultraspherical polynomials with rapidly growing tails.  In contrast, the proposed TDIE method utilizes orthogonal spherical harmonics instead of wave functions as the basis functions. A novel use of a time domain Green's function expansion in terms of spherical harmonics avoids convolutions with Ultraspherical polynomials while still allowing mode-by-mode solution (orthogonality is still retained). Though constructed within a integral equation (IE) based numerical solution framework, the proposed method does not suffer from geometry representation error and singular integrals, which are common for other IE-based methods. This formulation circumvents the problem of nonexistence of the inverse Fourier transform of the Hankel function, giving a rigorous and generalized framework for reconstructing the field on the surface. The resulting method exhibits excellent stability and accuracy even for very long time simulations.

The principle contributions of this work are: (1) A mesh-free TDIE framework with tesseral harmonics as spatial source and testing basis sets and a band limited temporal basis set. (2) A spherical expansion of the time domain Green's function that allows closed form evaluation of 4-dimensional spatial integrals as well as efficient numerical quadrature between smooth, bounded functions in computing temporal convolutions.  (3) A linear complexity Marching on in Time scheme with excellent late time stability that is accurate when compared with frequency domain Mie theory. (4) A series of results demonstrating the high order accuracy and convergence of the method for sound-soft, sound-rigid and penetrable spheres.  Note that in this work, we restrict the pulse excitation to be a plane wave; however, the same technique may be used for any band limited incident beam form.

The remainder of the paper is organized as follows. In Section \ref{sec:prob_statement} the time domain integral equations are formulated to describe the boundary value problem, with specializations to the soft, rigid and penetrable cases.  Exact equivalence between the Mie solution and the frequency domain version of the integral solution is demonstrated in Section \ref{sec:FDIE}.  Section \ref{sec:TDGF} provides the expansion of the time domain Green's function in spherical coordinates that underlies the formulation of the TDIE-MOT system in Section \ref{sec:TDIE}.  Section \ref{sec:res} presents results that validate the accuracy, convergence, and stability of the method.  Finally, we conclude and give future directions in Section \ref{sec:conc}.

\section{Problem Statement\label{sec:prob_statement}}

Consider a spherical scattering object that occupies a volume $D_1 \subset \mathbb{R}^3$ residing in a homogeneous background medium occupying $D_0=\mathbb{R}^3$. The boundary of the scatterer is denoted using $\Omega_1 = \partial D_1$, and is equipped with an outward pointing normal $\hat{n} (\br)$.  Assume that the media in both $D_0$ and $D_1$ are irrotational, inviscid, and slightly compressible so that the velocity $\vv{v}(\vv{r},t)$ in each region may be characterized via the relation $\vv{v}_q(\vv{r},t)=\nabla \varphi_q(\vv{r},t)$, $q=0,1$, where $\varphi_q(\br,t)$ is the velocity potential in region $D_q$.  The constitutive parameters of the background and the object are respectively denoted $\left \{ \rho_0, \nu_0\right \} $, and $\left \{ \rho_1, \nu_1\right \}$, where $\rho_q$ and $\nu_q$ for $q = 0,1$ are the density and speed of sound in each region $D_q$. Assuming that the object is at rest with respect to the background medium, the pressure may be obtained as $p_q(\br,t)=\rho_q\partial_t \varphi_q(\br,t)$, 
where $\partial_t$ denotes derivative with respect to time.  An acoustic pulse characterized by $\varphi^i(\br,t)$ which contains information of $\left \{ p^i(\br,t), \vv{v}^i(\br,t) \right \}$ that is band limited to a maximum frequency of $f_{max}$ and vanishingly small for $t<0$, is incident on the scatterer. The total velocity potential for $\br \in D_0$ comprises both the incident and in scattered field, viz., $\varphi_0 (\br,t) = \varphi_0^i (\br,t) + \varphi_0^s (\br,t)$ whereas the velocity potential for $\br \in D_1$ is purely due to the scattered field: $\varphi_1(\br,t)=\varphi^s_1(\br,t)$. Boundary conditions satisfied by the velocity potential follow those satisfied by the pressure and velocity, and are as follows:
\begin{subequations}\label{equ:bc}
\begin{equation} \label{equ:BC1} 
\rho_0\partial_t \varphi_0(\bar{r}, t) = \rho_1\partial_t \varphi_1 (\bar{r}, t) ~~\forall \bar{r} \in \Omega_1
\end{equation}
\begin{equation} \label{equ:BC2} 
\dfrac {\partial \varphi_0 (\bar{r}, t) }{\partial n}=\dfrac {\partial \varphi_1 (\bar{r}, t) }{\partial n} ~~\forall \bar{r} \in \Omega_1
\end{equation}
\end{subequations}
Using the Kirchhoff-Helmholtz theorem and \eqref{equ:bc}, the velocity potential satisfies the following equations for cases $ \bar {r} \in \Omega_1^+ $ (when $\bar{r}$ approaches $\Omega_1$ from $D_0$) and $\bar{r} \in \Omega_1^-$ (when $\bar{r}$ approaches $\Omega_1$ from $D_1$) respectively
\begin{subequations}
  \begin{equation} \label{equ:TD1}
    \begin{split}
		 \partial_t \varphi^i(\bar{r},t) + \partial_t \int_{\Omega_1^{+}}\left[\varphi_0(\bar{r}',t )\star \dfrac{\partial G_0(\bar{r},\bar{r}',t)}{\partial n'} 
  -G_0(\bar{r},\bar{r}',t)\star \dfrac{\partial \varphi_0(\bar{r}',t)}{\partial n'} \right]dS'
  = \partial_t \varphi_0(\bar{r},t)
 \end{split}
 \end{equation}
\begin{equation}\label{equ:TD2}
       \begin{split}
			 \partial_t\int_{\Omega_1^{-}}\left[ \varphi_1(\bar{r}',t )\star \dfrac{\partial G_1(\bar{r},\bar{r}',t)}{\partial n'} 
-G_1(\bar{r},\bar{r}',t)\star\dfrac{\partial \varphi_1(\bar{r}',t)}{\partial n'}\right]dS'
 =- \partial_t \varphi_1(\bar{r},t)
  \end{split}
 \end{equation}
\end{subequations}
where $\partial_t$ denotes a derivative w.r.t time, ``$\star$'' denotes temporal convolution, $G_q(\bar{r},\bar{r}',t) = \dfrac{1}{4 \pi} \dfrac{\delta \left ( t - \left | \bar{R} \right |/v_q\right )}{\left | \bar{R} \right |}$, and $\bar{R} = \bar{r} - \bar{r}'$.  The temporal derivative is taken on both sides of the equation to facilitate application of the boundary conditions \eqref{equ:bc}, which occurs as the next step of the formulation.  Doing so may be interpreted as using the more physical quantities of pressure and its normal derivative as unknowns instead of the corresponding velocity potential quantities.  To simplify notation, the following time-dependent single-layer and double layer potential operators are defined.
\begin{subequations}
 \begin{equation}
 {\cal S}_q(\varphi)  = \int_{\Omega} G_q(\bar{r},\bar{r}',t)\star\dfrac{\partial \varphi(\bar{r}',t)}{\partial n'}dS'
\end{equation}
 \begin{equation}
 {\cal D}_q(\varphi)  = \int_{\Omega} \varphi(\bar{r}',t )\star \dfrac{\partial G_q(\bar{r},\bar{r}',t)}{\partial n'} dS'
\end{equation}
\end{subequations}

Using \eqref{equ:bc} to rewrite \eqref{equ:TD2}, the limiting cases (where both $\bar{r} \in \Omega_1$ and $\bar{r'} \in \Omega_1$) of \eqref{equ:TD1} and \eqref{equ:TD2} can be written as 
\begin{subequations}\label{equ:TDlim}
  \begin{equation} \label{equ:TD1lim}
    \begin{split}
 \partial_t \varphi^i(\bar{r},t) + \partial_t {\tilde {\cal D}}_0(\varphi_0(\bar{r}',t))  -\partial_t  {\cal S}_0(\varphi_0(\bar{r}',t))
  = \frac{1}{2}\partial_t \varphi_0(\bar{r},t)
 \end{split}
 \end{equation}
  \begin{equation}\label{equ:TD2lim}
   \partial_t \dfrac {\rho_0}{\rho_1}{\tilde {\cal D}}_1(\varphi_0(\bar{r}',t)) 
- \partial_t  {\cal S}_1(\varphi_0(\bar{r}',t))
 =-\dfrac {\rho_0}{2 \rho_1}\partial_t\varphi_0(\bar{r},t)
 \end{equation}
 \end{subequations}
where ${\tilde {\cal D}}_q$ means taking the integral in ${\cal D}_q$ in the Cauchy Principle Value (CPV) sense. Equations \eqref{equ:TD1lim} and \eqref{equ:TD2lim} for $\bar{r} \in \Omega_1$ form the cornerstone of our analysis. These equations apply to the penetrable scatterer case;  reduced forms  for the sound-soft and sound-rigid cases are obtained by imposing the usual modified boundary conditions.

For sound-soft sphere, pressure on the surface vanishes ($\partial_t \varphi_0 = 0$), therefore only one IE is left to represent the scattering problem.

\begin{equation} \label{equ:SoundSoft}
    \begin{split}
 \partial_t \varphi^i(\bar{r},t)  -\partial_t  {\cal S}_0(\varphi_0(\bar{r}',t)) 	 = 0
 \end{split}
 \end{equation} 
Similarly for sound-rigid case, the normal velocity on the surface vanishes ($\partial_n \varphi_0 = 0$), and the corresponding IE is written as
  \begin{equation} \label{equ:SoundRigid}
    \begin{split}
 \partial_t \varphi^i(\bar{r},t) + \partial_t {\tilde {\cal D}}_0(\varphi_0(\bar{r}',t)) 
  = \dfrac{1}{2}\partial_t \varphi_0(\bar{r},t)
 \end{split}
 \end{equation}

Solution to these integral equations proceeds via two stages: (i) representation of the geometry and (ii) representation of potentials that reside on this geometry. The first is trivial: the spherical form of our structure permits an exact geometric representation. What is left therefore, is to represent the potential on the surface of the object. To this end, we choose to represent the potential using a basis of tesseral harmonics. In what follows, for the sake of completeness, we shall review using these basis functions in the frequency domain, and illustrate the close relationship between integral equations discretized using tesseral harmonics and analytical Mie solutions. This will be then followed by deriving analogous expressions in time domain. 

\section{Equivalence between frequency domain Mie and IE solutions\label{sec:FDIE}}

In what follows, we briefly digress to draw equivalences between the time-harmonic version of the integral equations \eqref{equ:TD1},\eqref{equ:TD2} and classical analytical solutions to these equations. In the Fourier domain, \eqref{equ:TD1} and \eqref{equ:TD2} can be written as 

\begin{subequations}\label{equ:FDIE}
 \begin{equation}\label {equ:FDIE1}
 \varphi^i(\bar{r}) + \int_{\Omega_1^+}\left[\varphi_0(\bar{r}' ) \dfrac{\partial G_0(\bar{r},\bar{r}')}{\partial n'} 
 -G_0(\bar{r},\bar{r}')\dfrac{\partial \varphi_0(\bar{r}')}{\partial n'} \right]dS'=\varphi_0(\bar{r})
 \end{equation}
  \begin{equation}\label {equ:FDIE2}
   \int_{\Omega_1^-}\left[ \varphi_1(\bar{r}' ) \dfrac{\partial G_1(\bar{r},\bar{r}')}{\partial n'} 
 -G_1(\bar{r},\bar{r}')\dfrac{\partial \varphi_1(\bar{r}')}{\partial n'} \right]dS'=- \varphi_1(\bar{r})
 \end{equation}
 \end{subequations}

and in the corresponding limiting cases as:

\begin{subequations}\label{equ:FDIElim}
 \begin{equation}\label {equ:FDIE1lim}
 \varphi^i(\bar{r}) + \dashint_{\Omega_1} \varphi_0(\bar{r}' ) \dfrac{\partial G_0(\bar{r},\bar{r}')}{\partial n'} dS'
 -\int_{\Omega_1} G_0(\bar{r},\bar{r}')\dfrac{\partial \varphi_0(\bar{r}')}{\partial n'} dS'=\dfrac{1}{2}\varphi_0(\bar{r})
 \end{equation}
  \begin{equation}\label {equ:FDIE2lim}
   \dashint_{\Omega_1} \dfrac {\rho_0}{\rho_1} \varphi_0(\bar{r}' ) \dfrac{\partial G_1(\bar{r},\bar{r}')}{\partial n'}dS' 
 -\int_{\Omega_1} G_1(\bar{r},\bar{r}')\dfrac{\partial \varphi_0(\bar{r}')}{\partial n'} dS'=- \dfrac {\rho_0}{2 \rho_1} \varphi_0(\bar{r})
 \end{equation}
 \end{subequations}
 where $G_q(\bar{r},\bar{r}')  = \exp[-j \omega |\bar{R}|/v_q]/(4\pi |\bar{R}|)$ is the frequency domain Helmholtz Green's function for $q = 0,1$  and an $e^{j\omega t}$ time dependence has been suppressed. The solution to these equations can be obtained assuming that the unknown potential can be represented using a space of basis functions. Here, the basis functions are assumed to be spherical harmonics, viz., 
\begin{eqnarray}\label{equ:basisFn}
	\varphi_0 (\bar{r}) & = & \sum_{n=0}^{N}\sum_{m = -n}^n c_{nm}  Y_{n}^m (\theta,\phi)\\
\partial_n \varphi_0 (\bar{r})& = & \sum_{n=0}^N\sum_{m = -n}^n d_{nm} Y_{n}^m (\theta,\phi) 
\end{eqnarray}
where
\begin{equation}
Y_n^m(\theta,\phi) = \sqrt{  \dfrac {2n+1}{4\pi}  \dfrac {(n-m)!}{(n+m)!}  }  P_n^m(\cos \theta)e^{jm\phi}
\end{equation}
are the othornormalized version of spherical harmonics and $N$ is the truncation limit of the summation. In what follows, $\sum_{n,m}$ is used to denote $\sum_{n=0}^{n=\infty}\sum_{m=-n}^{m=n}$.
The addition theorem for the frequency domain Green's function and representation of the incident plane wave in terms of spherical harmonics are given as
\begin{equation}\label{equ:FDAddThrm}
 \dfrac{e^{-jk_q|\bar{r}-\bar{r}'|}}{4\pi |\bar{r}-\bar{r}'|} 
  =  -jk\sum_{n,m}  h_{n}^{(2)}(k_q |\bar{r}|_>) j_{n}(k_q |\bar{r}|_<) Y_n^m(\theta_>,\phi_>){Y_n^m}^\ast(\theta_<,\phi_<) 
\end{equation}
\begin{equation}\label{equ:plane2}
 e^{-j k_0 \hat{k}_0^i \cdot \bar{r}} =  \sum_{n,m} 4 \pi j^{-n} j_{n}(k_0 |\bar{r}|) Y_{n}^{m}(\theta,\phi){Y_{n}^{m}}^\ast(\theta^i,\phi^i)
\end{equation}
In the above expressions, $j_{n}(\cdot)$ is an $n$th order spherical Bessel function, $h_{n}^{(2)}(\cdot)$ is an $n$th order spherical Hankel function of the second kind, and $k_q=\omega/\nu_q$ is the wave constant in medium $q$. We note that the expansion for the Green's function is convergent only for $|\br|\gtrless|\br'|$.  As a result, \eqref{equ:FDAddThrm} must be used with the non-limiting cases of the Fourier domain integral equations \eqref{equ:FDIE}.  After discretization and testing with the truncated tesseral harmonic expansions \eqref{equ:basisFn}, orthogonality of the basis sets with the expansion \eqref{equ:FDAddThrm} results in a finite sum, and the limits $\Omega_1^+\rightarrow\Omega_1$ and $\Omega_1^-\rightarrow\Omega_1$ may be taken without issue.   Using expressions \eqref{equ:FDAddThrm}, \eqref{equ:plane2} together with the definitions of basis functions \eqref{equ:basisFn} in \eqref{equ:FDIE}, taking the inner product of the resulting system with $Y_n^{m*}(\theta,\phi)$ results a system of equations for each mode $nm$ (after the limiting process):
\begin{equation}\label{mat:fd}
\begin{split}
 \begin{bmatrix}
  A_0  &  
   4\pi  a^2 + B_0
  \\ A_1  & 
     \frac{\rho_0}{\rho_1}  [-4\pi  a^2 + B_1]
 \end{bmatrix}
 \begin{bmatrix}
  d_{nm}
  \\ c_{nm}
 \end{bmatrix}
= 
 \begin{bmatrix}
  F  \\ 0
 \end{bmatrix}
\end{split}
\end{equation}
where each of the elements are given as follows: 
\begin{equation} \label{ABCD:11}
 A_0 = -4\pi a^4 jk_0 h_{n}^{(2)}(k_0 a) j_{n}(k_0 a)  
\end{equation}
\begin{equation}\label{ABCD:21}
 A_1 = -4\pi a^4 jk_1 h_{n}^{(2)}(k_1 a) j_{n}(k_1 a) 
\end{equation}
\begin{equation}\label{ABCD:31}
 B_0 = 4\pi a^4 jk_0^2 h_{n}^{(2)}(k_0 a) j_{n}'(k_0 a) 
\end{equation}
\begin{equation}\label{ABCD:41}
 B_1 = 4\pi a^4 jk_1^2 h_{n}^{'(2)}(k_1 a) j_{n}(k_1 a) 
\end{equation}
\begin{equation}\label{ABCD:51}
F =(4\pi)^2 a^2 j^{-n}{Y_{n}^{m}}^\ast(\theta_k,\phi_k)j_{n}(k_0 a)
\end{equation}
The above expressions are obtained in the limit that $|\bar{r}| = |\bar{r}|' = a$. 

Solution to \eqref{mat:fd} yields coefficients for the total velocity potential and its normal derivative
\begin{subequations}\label{equ:FDcoeffs}
\begin{equation} \label{form:phi_1}
\begin{split}
c_{nm} = \dfrac{4 \pi j^{n-1} \rho_1 j_n(k_1 a) {Y_{n}^{m}}^\ast(\theta^i,\phi^i)   }
     {a^2 k_0 [k_0 \rho_1 j_n(k_1 a) h_n^{'(2)}(k_0 a) - k_1 \rho_0 j_n'(k_1 a) h_n^{(2)}(k_0 a)  ] }
\end{split}
\end{equation}
\begin{equation}\label{form:dphi_1}
\begin{split}
d_{nm} 
= \dfrac{4 \pi j^{n-1} \rho_0 k_1 j_n'(k_1 a) {Y_{n}^{m}}^\ast(\theta^i,\phi^i)  }
     {a^2 k_0 [k_0 \rho_1 j_n(k_1 a) h_n^{'(2)}(k_0 a) - k_1 \rho_0 j_n'(k_1 a) h_n^{(2)}(k_0 a)  ] }
\end{split}
\end{equation}
\end{subequations}
As opposed to the above approach, analytical Mie solution approaches start by representing the scattered velocity potential in terms of outgoing spherical waves, and those inside using incoming spherical waves. The coefficients are obtained using boundary conditions and orthogonality. It can be readily verified that the coefficients of the spherical harmonics used to represent the velocity potential in a traditional Mie approach are identical to those obtained in \eqref{equ:FDcoeffs}. Furthermore, it can be shown that the integral equations \eqref{equ:FDIE} are unique and not corrupted by internal resonances.  

One can observe similar equivalence between IE and Mie solutions for sound-soft and sound-rigid spheres, which means the analytical result for the IE should lead to Mie solution.  However, when analyzing either purely rigid or soft cases using integral equations, it is well known that discretization of these equations results in a non-trivial projection onto the null-space of the operators (in the analytical scenario, the null-space is avoided through cancellation by the numerator). For the discretized problem, various methods have been developed to overcome this problem, including Burton-Miller \citep{Burton1971} and CHIEF \citep{Schenck1968}. 

While the above exposition dealt with frequency domain, obtaining the transient response is not as trivial as taking the inverse Fourier transform of \eqref{equ:FDcoeffs} because the inverse Fourier transform of the spherical Hankel function is not defined. In what follows, we present an approach using a novel representation of the retarded potential Green's function in \eqref{equ:TD1lim} and \eqref{equ:TD2lim} together with \eqref{equ:basisFn} to create an MOT framework to compute transient coefficients associated with interior and exterior velocity potentials. 

\section {Time-Domain Mie Solution}

This section presents an approach for deriving a time domain analog of the integral-equation based Mie solutions presented in the previous section. The starting point of this analysis is the development of a novel expansion for the retarded potential Green's function in terms of spherical harmonics without taking recourse to the addition theorem presented in \eqref{equ:FDAddThrm}. 

\subsection{Spherical Expansion of Time-Domain Green's Function\label{sec:TDGF}}

Consider any function $f(x)$ that can be expanded by Legendre polynomials.
\begin{equation}
 f(x) = \sum_{n=0}^\infty a_n P_n(x);~~ a_n=\frac{2n + 1}{2} \int_{-1} ^ {1} f(x) P_n(x) dx
\end{equation}
Suppose there exist coordinates denoted as $\bar{r}_1$, $\bar{r}_2$, with $r_1=|\br_1|$, $r_2=|\br_2|$ and $R=|\bar r_1-\bar r_2|$. One can then define 
\begin{equation}
x = \cos \gamma = \frac {r_1^2 + r_2^2 -R^2} {2 r_1 r_2}
\end{equation}
The coefficients of the Legendre expansion may then be rewritten as 
\begin{equation} \label{leg:coeff}
\begin{split} 
 a_n = \dfrac{2n + 1}{2 r_1 r_2} \int_{|r_1 - r_2|} ^ {r_1 + r_2} f\left (\frac {r_1^2 + r_2^2 -R^2} {2 r_1 r_2} \right ) 
           P_n \left (\frac {r_1^2 + r_2^2 -R^2} {2 r_1 r_2} \right ) R dR
\end{split}
\end{equation}
Now, let $f(x)$ denote the unbounded Green's function in a medium with wave number $k_q=\omega/v_q$,
\begin{equation}
 f(x) = f \left (\frac {r_1^2 + r_2^2 -R^2} {2 r_1 r_2} \right ) =  \dfrac {e^{-jk_q R}} {4 \pi R}
\end{equation}
The Green's function may now be represented in terms of Legendre polynomials can be written as 
\begin{equation}\label{equ:FDleg}
  \dfrac {e^{-jk_q R}} {4\pi R} = \sum_{n=0}^\infty \dfrac{2n + 1}{8 \pi r_1 r_2} P_n \left ( \cos \gamma \right ) \int_{|r_1 - r_2|} ^ {r_1 + r_2} e^{-jk_q R}
           P_n\left (\frac {r_1^2 + r_2^2 -R^2} {2 r_1 r_2} \right )dR 
\end{equation}
An Inverse Fourier transform (IFT) of both sides of \eqref{equ:FDleg} with respect to $\omega$ yields the desired representation of the retarded potential Green's function 
\begin{equation}
\begin{split} \label{td:exp}
  \dfrac {\delta \left (t- \frac{R}{v_q}\right )} {4\pi  R}=\sum_n \dfrac{(2n + 1)v_q}{8\pi  r_1 r_2} 
           P_n\left (\frac {r_1^2 + r_2^2 -v_q^2 t^2} {2 r_1 r_2} \right ) \mathcal{P}_{\alpha \beta}(t) P_n( \cos \gamma )
 \end{split}
\end{equation}
where $\mathcal{P}_{\alpha \beta}(t) $ is a pulse function (the value is $1$ when $t \in [\alpha,\beta]$, and zero otherwise) with $\alpha = |r_1 - r_2|/v_q $ and $ \beta =(r_1 + r_2)/ {v_q}$. 
Finally, by employing the addition theorem for Legendre polynomials, one can obtain an expansion for the time domain Green's function in spherical coordinates
\begin{equation}
\begin{split} \label{td:exp1}
  \dfrac {\delta \left (t- \frac{R}{v_q} \right )} {4\pi  R}=\sum_{nm} \dfrac{(2n + 1)v_q}{8\pi  r_1 r_2} 
           P_n\left (\frac {r_1^2 + r_2^2 -v_q^2 t^2} {2 r_1 r_2} \right ) \mathcal{P}_{\alpha \beta}(t) Y_n^m(\theta_1,\phi_1){Y_n^m}^\ast(\theta_2,\phi_2)
 \end{split}
\end{equation}
where $\theta_i$ and $\phi_i$ are polar and azimuthal angles corresponding to $\bar{r}_i$ for $i = 1,2$. When $r_1=r_2=a$ (i.e. $\br_1,\br_2\in\Omega$ with the origin at the sphere center), equation \eqref{td:exp1} separates spatial and temporal dependence of the time domain Green's function.  Furthermore, because tesseral harmonics are by definition separable in their arguments, \eqref{td:exp1} is separable in all arguments when restricted to the sphere surface.  As is demonstrated in the next section, this allows closed form evaluation of all spatial integrals in the TDIE system matrix. A similar expression was provided in the literature \citep{Buyukdura1997}, {\em sans} the above derivation or its use in analysis. In what follows, this expression is used to construct a marching-on-in-time scheme to solve for scattering from soft, rigid and penetrable objects in time domain. 

\subsection{Discretization of TDIE} \label{sec:TDIE}

The first step in solving \eqref{equ:TD1lim} and \eqref{equ:TD2lim} using the representation of the Green's function given in Section \ref{sec:TDGF} is to represent the velocity potentials (and their normal derivatives) in terms of spatial and temporal basis functions. To this end, assume that the potentials can be represented as
\begin{subequations}\label{equ:tdSpBasis}
\begin{eqnarray}
\varphi_0 (\bar{r},t)  =  \sum_{n=0}^N\sum_{m = -n}^n  c_{nm}(t) Y_{n}^m (\theta,\phi)
 =  \sum_{n=0}^N\sum_{m = -n}^n \sum_{k=1}^{N_t} c_{k,nm}  T_k(t) Y_{n}^m (\theta,\phi)\\
\partial_n \varphi_0 (\bar{r},t) =  \sum_{n=0}^N\sum_{m = -n}^n  d_{nm} (t) Y_{n}^m (\theta,\phi) 
 = \sum_{n=0}^N\sum_{m = -n}^n \sum_{k=1}^{N_t} d_{k,nm} T_k(t) Y_{n}^m (\theta,\phi) 
\end{eqnarray}
\end{subequations}
where $ c_{nm}(t)$ and $ d_{nm}(t)$ are the transient signatures of each spatial mode, and $c_{k,nm}$ and $d_{k,nm}$ are the final unknown coefficients to be solved for.
In the above expansions $T_k (t) = T(t - k\Delta_t)$ are the temporal basis functions, $\Delta_t$ is the time step size, and $N$ truncates the spatial basis expansions. While there are several choices for temporal basis functions including backward looking Lagrange polynomials \citep{Ergin1998}, and splines \citep{Davies2013}, in this paper we use approximate prolate spheroidal wave functions (APSWFs) as interpolants \citep{Knab1979}. These functions are strictly band limited and approximately time limited functions and have been shown to have excellent interpolatory properties \citep{Knab1979}, and have found widespread use in time domain integral equation analysis \citep{Ergin2000,Pray2012}. They are given as 
\begin{equation}
T_k(t)=\dfrac{\omega_o}{\omega_s}
\dfrac{\sin(\omega_o(t-k\Delta_t))}{\omega_o(t-k\Delta t)}
\dfrac{\sinh(\frac{\pi}{2}p_t(1-1/\chi_o)\sqrt{1-[(t-k\Delta t)/p_t\Delta_t]^2})}
{\sinh(\frac{\pi}{2}p_t(1-1/\chi_o)) \sqrt{1-[(t-k\Delta t)/p_t\Delta_t]^2}}
\end{equation}
where $\chi_o$ is the oversampling ratio, $\omega_s=\frac{\pi}{\Delta_t}=\chi_o \omega_{max}$,  $\omega_o=\frac{\omega_s+\omega_{max}}{2}$, and $p_t$ defines the half-width of the function.  The APSWF $T_k(t)$ is symmetric about $t=k\Delta t$, and therefore possesses a non-causal portion of duration $p_t$.

Using \eqref{equ:tdSpBasis} in \eqref{equ:TD1lim} and \eqref{equ:TD2lim}, testing in space with $Y_n^{m\ast} (\theta,\phi)$ and in time with a delta function $\delta (t)$ results in a system of equations that may be compactly represented as 
\begin{equation}\label{equ:preMOT}
\sum_{i=0}^{p_t} {\cal Z}_{-i} {\cal I}_ {j +i}= {\cal F}_j - \sum_{i= max \{j-k_{max},1\} } ^{j-1} {\cal Z}_{j-i} {\cal I}_{i}
\end{equation}
where $k_{max}=[\frac{R}{v\Delta_t}]+p_t$. 
Explicit expression for matrices in \eqref{equ:preMOT} will be given in the subsequent sub-sections. Recall that the system of equations is not causal thanks to the choice of the temporal basis function. Methods to convert this equation to a casual system using band limited extrapolation have been prescribed \citep{Wildman2004}, and are used here. The resulting set of equations can be written as 
\begin{equation}\label{equ:MOT}
 {\cal \tilde Z}_{0} {\cal I}_ {j} = {\cal F}_j - \sum_{i= max \{j-k_{max},1\} } ^{j-1} {\cal \tilde Z}_{j-i} {\cal I}_{i} 
\end{equation}
where ${\cal \tilde Z}_{k}$ for all $k$ are related to ${\cal Z}_k$ and can be obtained from the expressions given by Wildman et al.\citep{Wildman2004}  This equation forms a marching-on-in-time system that can be used to recursively obtain the time signatures, $c_{nm} (t)$ and $d_{nm} (t)$, of the coefficients of the spherical harmonics. In the above equations, the forcing function ${\cal F}$ is defined as
\begin{equation}
{\cal F}_{nm,i} = \left . \int_{\Omega_1} dS' Y_n^{m\ast} (\theta, \phi) \partial_t \varphi^i(\bar{r},t) \right |_{t = i\Delta_t}
\end{equation}
Next, we prescribe the elements of ${\cal Z}_k$ for three different equations: the cases of sound soft, hard and penetrable objects, respectively. 

\subsubsection{Sound Soft Object}

As is evident from \eqref{equ:SoundSoft}, the unknowns are those associated with the normal derivative of the velocity potential, $d_{nm}(t)$, i.e., ${\cal I}_{nm,i} = d_{i,nm}$. The elements of the ${\cal Z}_i$ can be obtained using 
\begin{equation}
\begin{split}
{\cal Z}_{nm,n'm', i}  & =
 \left .  \partial_t \int_{\Omega_1} {Y_n^m}^{*}(\theta, \phi) \int_{\Omega_1} \dfrac{\delta(t-\frac{R}{v_0}) }{4 \pi R} \star T(t) Y_{n'}^{m'}(\theta', \phi') dS'dS \right | _{t=i\Delta_t}\\
 & =   \left . \dfrac{v_0}{8\pi rr'} 
           P_n\left (\frac {r^2 + r'^2 -v_0^2 t^2} {2 rr'}  \right ) \mathcal{P}_{\alpha \beta}(t) \star\partial_t  T(t) \delta_{nn'}\delta_{mm'} 
\right | _{\substack{r=a \\ r'=a \\ t=i\Delta_t}}\\
  & \doteq {\cal L}\{i,n,v_0\}\delta_{nn'}\delta_{mm'}
 \end{split}
\end{equation}
where $\delta_{pp'}$, for $p =n,m$ are the Kronecker delta function. In the above expressions, we have used \eqref{td:exp1} and the basis set for $\partial_{n}\psi(\br,t)$ from \eqref{equ:tdSpBasis}. The orthogonality of the spatial basis functions renders the spatial integral trivial to evaluate leaving only temporal convolution to be numerically integrated. Note that since the temporal basis is strictly band limited, so is its convolution with the Legendre polynomial. Furthermore, the Green's function is a polynomial of order $2n$. As a result of this smoothness, this convolution can be evaluated efficiently using numerical quadrature. 

\subsubsection{Sound hard Object}

The sound hard case is slightly more complex as it involves the normal derivative of the Green's function, with the velocity potential as the unknown, i.e., ${\cal I}_{i,nm} = c_{i,nm}$. The elements of ${\cal Z}_i$ can be obtained using
\begin{equation}
\begin{split}
{\cal Z}_{nm,n'm',i} &  = \left . \Bigg[
 \int_{\Omega_1} {Y_n^m}^{*}(\theta, \phi) \partial_t T(t) Y_n^m(\theta, \phi) dS \right. \\ 
 & + \left. \int_{\Omega_1} {Y_n^m}^{*}(\theta, \phi) \int_{\Omega_1} \dfrac{\partial }{\partial n'}  \dfrac{\delta\left (t-\frac{R}{v_0} \right ) }{4 \pi R} \star \partial_t T(t) Y_n^m(\theta', \phi') dS' dS \Bigg]  \right |_{t=i\Delta_t}
\\
& =\left [\delta(t) +  \dfrac{\partial }{\partial n'} \left \{ \dfrac{ v_0}{2 rr'} 
           P_n\left (\frac {r^2  + r'^2-v_0^2 t^2} {2 rr'} \right )  \mathcal{P}_{\alpha \beta}(t)  \right \} \right] \star \partial_t T(t) \delta_{nn'}\delta_{mm'}  \Biggl| _{\substack{r=a \\ r'=a \\ t=i\Delta_t}}\\
& \doteq  \Big[\partial_t T(t) + {\cal K}\{i,n,v_0\} \Big ] \delta_{nn'}\delta_{mm'}
 \end{split}
\end{equation}
As before, the only integrals to be numerically evaluated are the temporal convolutions between smooth functions. 

\subsubsection{Penetrable Object}

Finally, the penetrable case requires both the velocity potential and its normal derivative as unknowns, i.e., ${\cal I}_{ i, nm} = \left \{c_{i,nm}, d_{i,nm} \right \}$. As a result, the ${\cal Z}_i$ combines all of the previously defined operators and can be rewritten as 
\begin{equation}
{\cal Z}_{nm, n'm', i} = \left [ \begin{array}{cc}
{\cal Z}_{nm,n'm',i}^{pp} & {\cal Z}_{nm,n'm',i}^{pv} \\
{\cal Z}_{nm,n'm',i}^{vp} & {\cal Z}_{nm,n'm',i}^{vv}
\end{array} \right ]
\end{equation}
where
\begin{subequations}
\begin{eqnarray}
{\cal Z}_{nm,n'm',i}^{pp}  = {\cal L}\{i,n,v_0\}\delta_{nn'}\delta_{mm'}
\end{eqnarray}
\begin{eqnarray}
{\cal Z}_{nm,n'm',i}^{pv}  = \Big[\partial_t T(t) + {\cal K}\{i,n,v_0\} \Big ] \delta_{nn'}\delta_{mm'}
\end{eqnarray}
\begin{eqnarray}
{\cal Z}_{nm,n'm',i}^{vp}  = {\cal L}\{i,n,v_1\}\delta_{nn'}\delta_{mm'}
\end{eqnarray}
\begin{eqnarray}
{\cal Z}_{nm,n'm',i}^{vv}  = \Big[\partial_t T(t) - {\cal K}\{i,n,v_1\} \Big ] \delta_{nn'}\delta_{mm'}
\end{eqnarray}
\end{subequations}

All of the time domain systems for the three different case can be solved efficiently using equation \eqref{equ:MOT}, whose complexity order is $\mathcal{O}(N_t)$ in time. For a spherical surface, the spatial system matrix is purely diagonal, which means, as in frequency domain, that the transient coefficients can be solved for one mode at a time. Therefore the TDIE solver is of linear complexity in both space and time. In next section numerical examples involving different scatterers are given that demonstrate the stability and accuracy of this time domain framework.

\section{Numerical Examples}\label{sec:res}

In this section, we provide results that illustrate the accuracy and stability of the method.  First, transient scattering results for sound-soft, sound-hard and penetrable spheres due to an incident pulse are presented. By taking a Fourier transform, the normalized spectra are validated against time domain Mie theory. Next, convergence of the method relative to temporal sampling and interpolation is studied by choosing different oversampling factors and temporal basis function widths.  Finally, a long time simulation is performed to demonstrate late time stability of the proposed solver. In this section, the problem studied is scattering due to a wide band modulated Gaussian pulse incident on a sphere of radius $10\mu$m. The normalized excitation pulse has a center frequency of $f_0=20$MHz, and a bandwidth of $B=15$MHz.  The sound speed in the exterior medium ($\br \in D_0$) is  $v_0=1500 $m/s, and the sound speed in the interior medium ($\br\in D_1$) is $v_1=340 $m/s. The density ratio between exterior and interior media is $1.2$. The time step size is chosen as $\Delta_t = \frac{1}{\chi_o f_{max}}$, where $f_{max} = f_0 + B$ and the ratio $\chi_o$ is related to the oversampling factor(ratio between sampling frequency $f_s$ and Nyquist frequency $f_N=2f_{max}$).  In the present tests the incident field propagates along the $\theta^i=0$\textdegree and $\phi^i=0$\textdegree direction, although it could be incident from any direction.

\subsection{Transient scattering}

First, we present scattering results for the sound-soft, sound-hard, and penetrable cases.  In each example, $\chi_o=20$ is chosen and $p_t = 5$ is used as the half-width of the temporal basis function.  Due to the size of the scattering sphere and the frequency content of the incident pulse, four modes are sufficient to give convergent results.  Note that, because the expressions derived for the time domain integral equations are equivalent to the Fourier transforms of the corresponding frequency domain expressions (as shown in section \ref{sec:FDIE}), the number of harmonics required to reconstruct the field is exactly the same as in a traditional frequency domain Mie solution, i.e.  $N_{mode}\sim ka$, with $k$ the wave number corresponding the maximum frequency.

First, the TDIE is applied to soft sphere case, to obtain the normal derivative of the velocity potential. 
Figure 1 plots the time domain normal derivative of velocity potential observed at one point ($\theta=0$\textdegree and $\phi=0$\textdegree) on the sphere surface compared with a result from the IFT of the frequency domain Mie solution, and Figure 2 shows the normalized frequency spectrum of the TDIE result, compared again with frequency domain Mie solution. From the frequency spectrum plot, it is clear that all of the frequency components within the band are accurately captured.

Next, the results for the sound-hard case are given in Figures 3 (transient velocity potential) and 4 (normalized frequency spectrum). The velocity potential corresponding to the incidence pulse is also shown in Figure 3 to illustrate the difference between the incident field and the total field including scattering field. A common practice in analyzing these types of problems is to assume that the total field is the same as the incident field.  The result in Figure 3 shows that, with the given problem parameters (which are within the ranges used in applications such as ultrasound contrast agent simulation), this approximation is not valid, and the total field must be solved self-consistently to obtain a correct solution.

The third example is the more complicated penetrable case in which both the velocity potential and its normal derivative are sought by solving the two coupled integral equations in Section \ref{sec:TDIE}. The total velocity potential and its frequency spectrum are given in Figures 5 and 6 respectively. For the penetrable case, resonance phenomena result in a response with duration much longer than that of the incident Gaussian pulse.  The corresponding resonance peaks are clearly visible in the frequency domain plot in Figure 6; all of these are captured very accurately by the proposed time domain simulation.  Note that the inverse Fourier transform of the time domain Mie solution is not given here to make the plot easily readable.  The accuracy of the time domain signal may be verified in the frequency domain plot of Figure 6.

\subsection{Accuracy and stability of TDIE solver}

The next examples investigate convergence of the TDIE method relative to the oversampling factor and the half-width $p_t$ of the APSWF.  The problem parameters are the same as for the previous three scattering results.  Results show convergence in the relative $l_2$ error norm for the frequency spectrum of the TDIE solver compared with the frequency domain Mie solution. Figure 7 shows three curves which represent relative $l_2$ error versus oversampling factor ($f_s/f_N$) for 0th, 4th and 8th mode.   Even for high order modes, the relative error decreases to approximately $10^{-7}$ with decrease in time step size. Another factor that influences the accuracy is the width ($2p_t$) of the temporal basis function. Error curves for the 0th and 4th mode versus $p_t$ are plotted in Fig. 8, where $\chi_o=20$ and $p_t$ is increased from $1$ to $6$.  By 
increasing the width of temporal basis function, the accuracy significantly improves due to the higher order interpolation capability of the APSWF.  There is a plateau in convergence at approximately $p_t=4$ that we attribute to other factors in the discretization and solution process, particularly in the inversion of the MOT system; the purpose of presenting this result is to show that the parameters for the APSWF used in the scattering results above are sufficiently converged with respect to $p_t$.
With both the validation results and the shown convergence performance, the proposed time domain solver matches well against the analytical frequency domain Mie Theory results.

In addition to accuracy, late-time stability is a very well known problem for TDIE solvers, and has been heavily studied by researchers in both the acoustics and electromagnetics communities \citep{Davies2013,Pray2012,HaDuong2003}.  Analyses in several of these works indicate that a major source of instability is error in numerically evaluating spatial integrals.  In the present work all spatial integrations are performed analytically and the temporal convolutions yield smooth functions. As a result the method exhibits excellent late-time stability. Figure 9 shows a stable solution at an observation point on the sphere for 200,000 time steps. 

\section{Conclusion}\label{sec:conc}

This paper presents a time domain boundary integral equation framework for analyzing acoustic scattering from spherical objects in a homogeneous medium.  The unknown velocity potential is spatially discretized in terms of tesseral harmonics, and temporally discretized in terms of strictly band limited Approximate Spheroidal Wave Functions.  An expansion of the time domain Green's function in terms of spherical functions avoids the necessity of taking an inverse Fourier transform of a spherical Hankel function, as was done in previous work, and permits separation of temporal and spatial integrations in the boundary integral operators when applied to a spherical geometry.  Furthermore, the Green's function expansion permits mesh free spatial discretization with analytical and singularity-free evaluation of spatial integrations in both the single and double layer potential operators.  Temporal convolutions involve band limited functions and therefore exhibit rapid convergence when numerically integrated.  Numerical results demonstrate that the global time domain matrix system is well behaved and the MOT process is stable, even for long simulation time.  Since the spatial basis functions are mutually orthogonal, the system matrices are diagonal, which leads to an algorithm that with cost that scales linearly in both storage and computational complexity. It has been demonstrated both analytically and numerically that the TDIE result converges to the traditional frequency domain Mie Theory solution, and is therefore capable of capturing all scattering phenomena without relying on simplifying assumptions.   

The method has several possible applications and extensions.  First, it provides direct insight into the transient scattering behavior of acoustically excitepd spherical scatterers, e.g. microbubbles for drug delivery.  Second, it can provide direct time domain validation for purely numerical methods.  Third, it may be extended to nonlinear problems, for instance those in which the bubble moves under the influence of the exciting acoustic field.  Fourth, it can be expanded to systems of multiple spherical scatterers using a time domain analog to the T-matrix method.  The last two topics in particular are the subject of future work.

\medskip
 \noindent \textbf{Acknowledgements}
 \setlength{\parindent}{0.7cm} 

The authors wish to acknowledge the support from NSF CCF-1018516 and NSF CMMI-1250261. D. Dault would like to acknowledge support from the NSF Graduate Research Fellowship Program.

 \newpage
 
 \begin{center}
 
 \large{Figure Captions}

 \end{center}

 \noindent
 Figure 1.  Transient field at observation point on sound-soft sphere 
 \smallskip

 \noindent
 Figure 2.  Frequency domain result for sound-soft sphere 
 \smallskip

 \noindent
 Figure 3.  Transient field  at observation point on sound-rigid sphere  
 \smallskip

 \noindent
 Figure 4.  Frequency domain result for sound-rigid sphere 
 \smallskip

 \noindent
 Figure 5.  Transient field  at observation point on penetrable sphere 
 \smallskip

 \noindent
 Figure 6.  Frequency domain result for penetrable sphere 
 \smallskip

 \noindent
 Figure 7.  Relative L2 error versus oversampling factor 
 \smallskip

 \noindent
 Figure 8.  Relative L2 error versus half width of temporal basis function 
 \smallskip

 \noindent
 Figure 9.  Stable results of long time MOT simulation 
 \smallskip

 
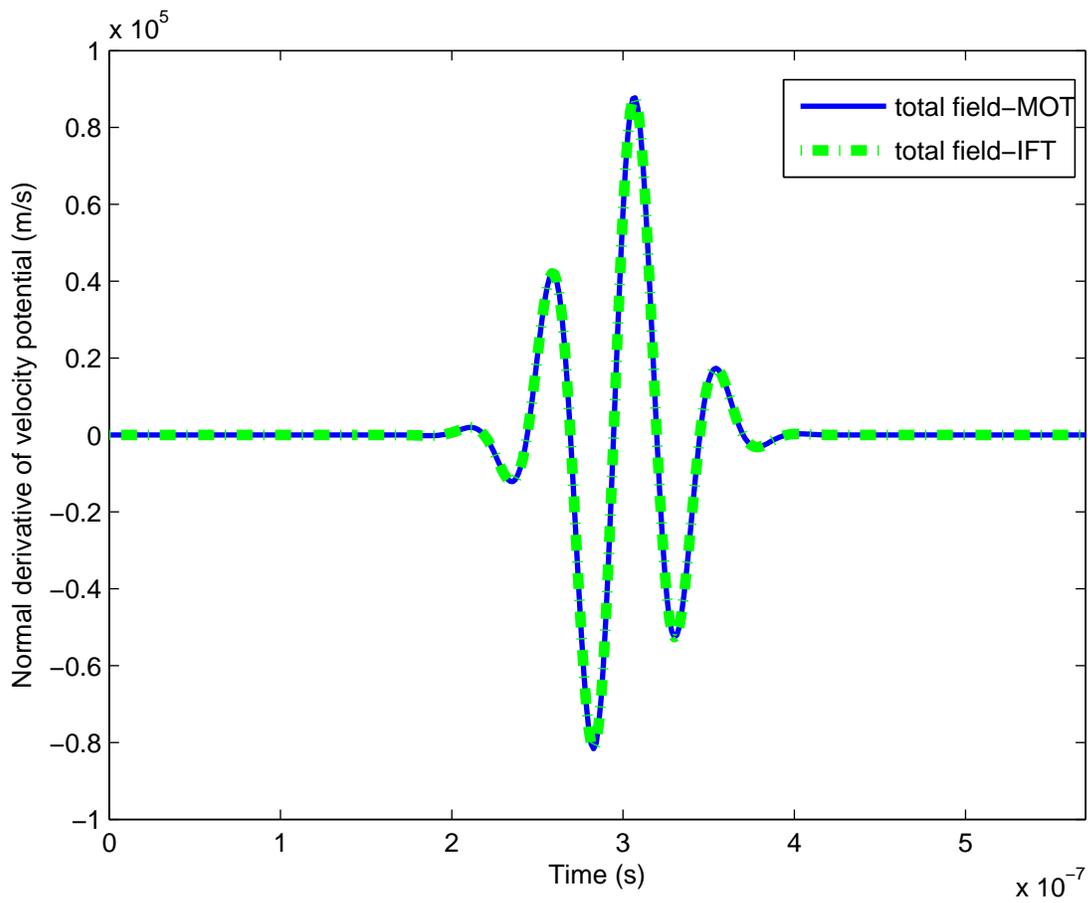
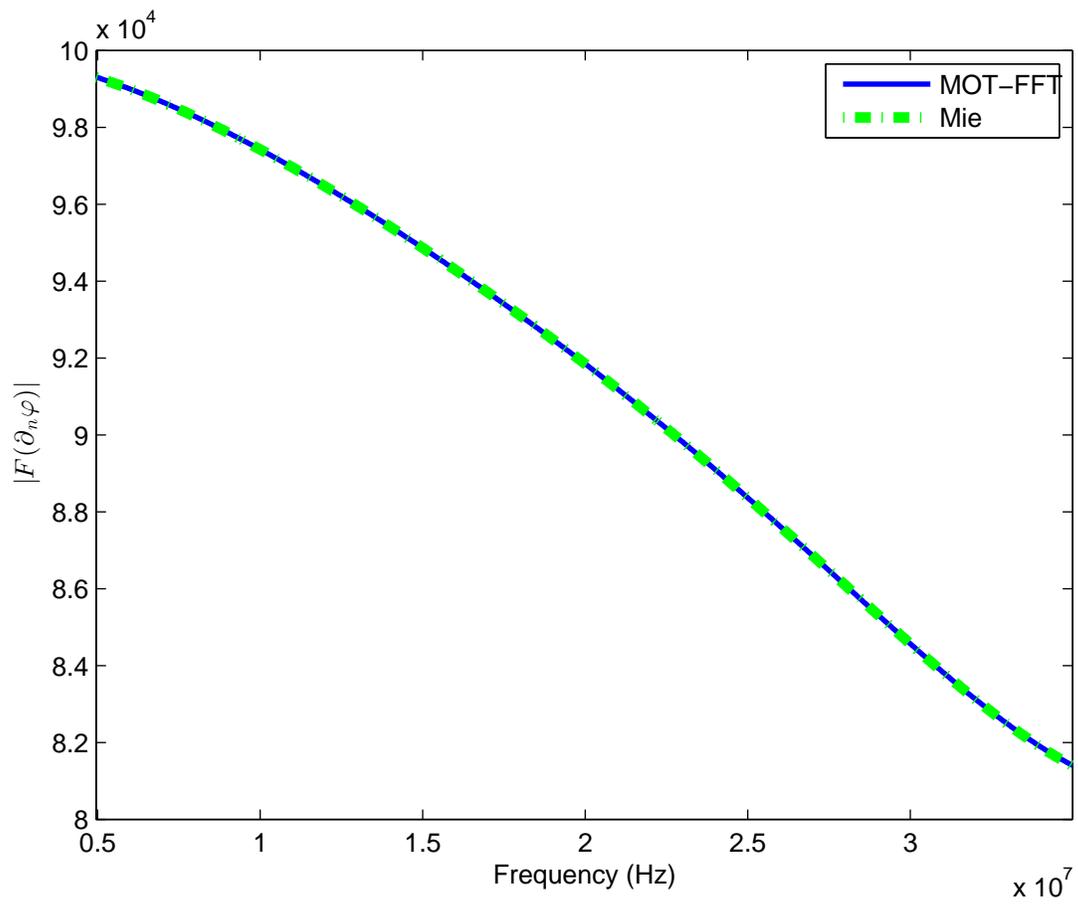
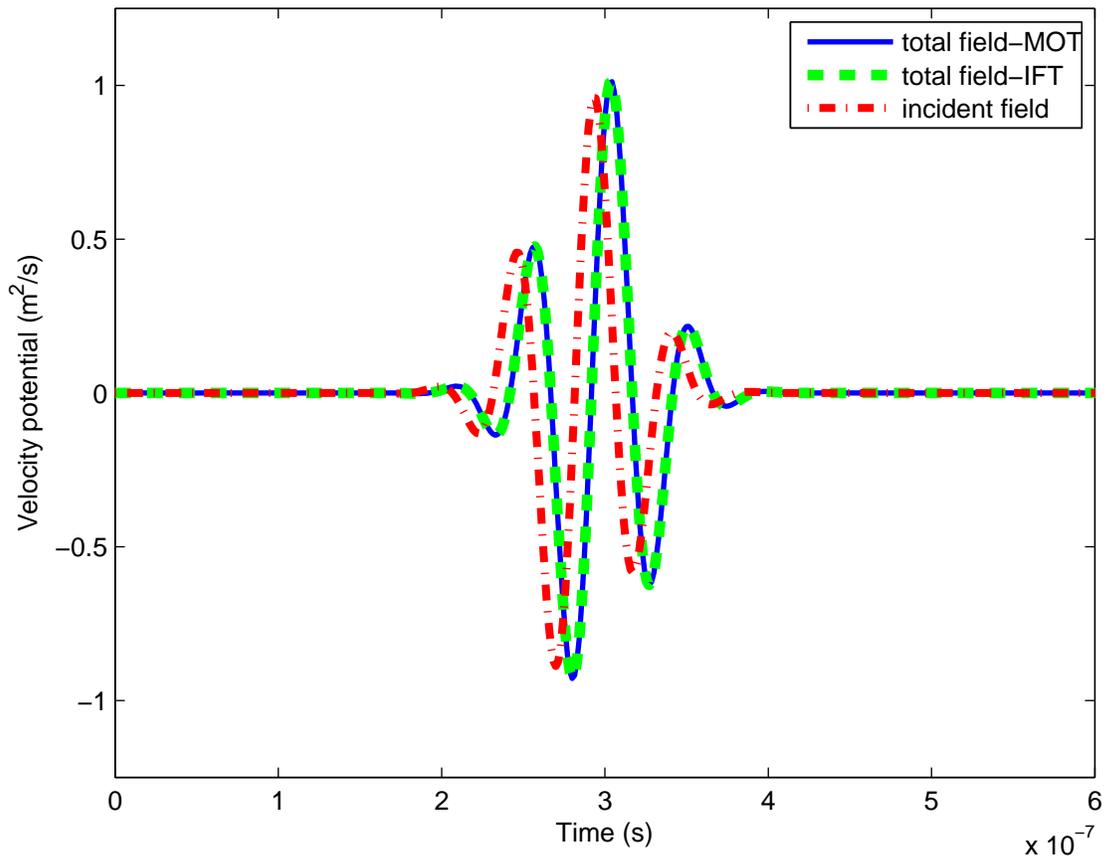
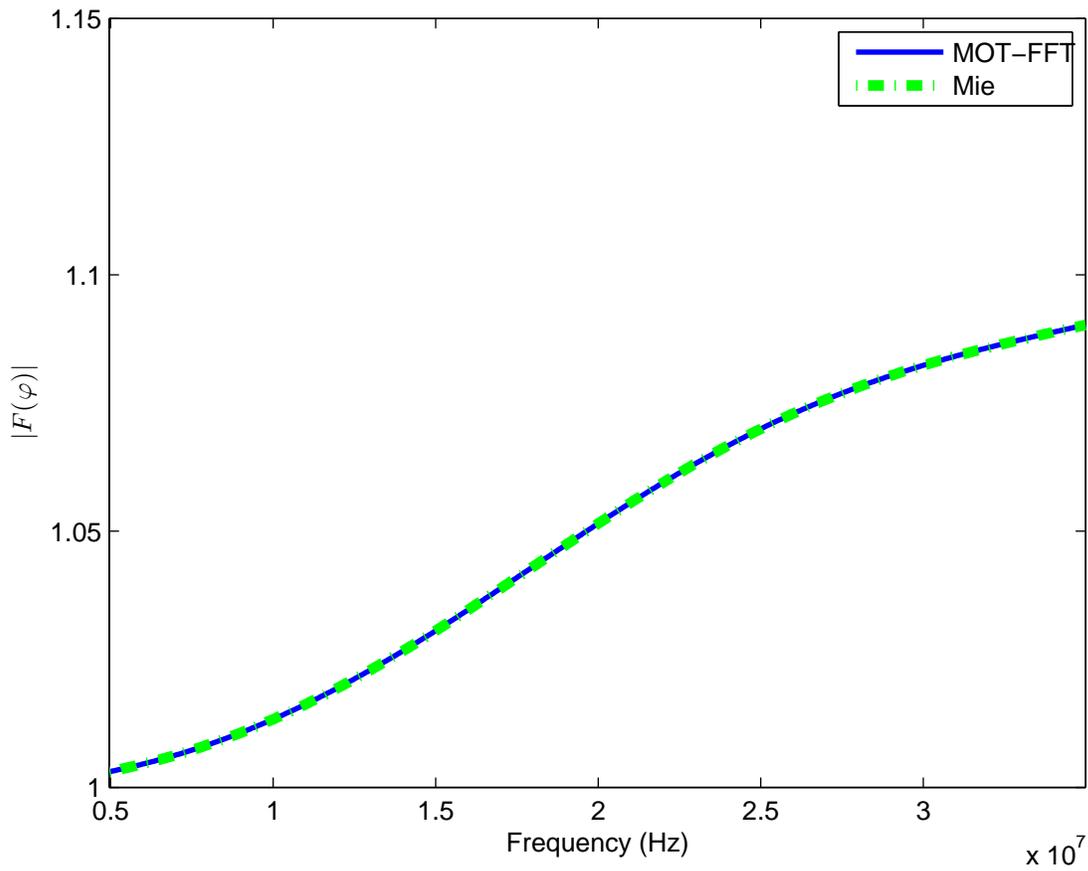
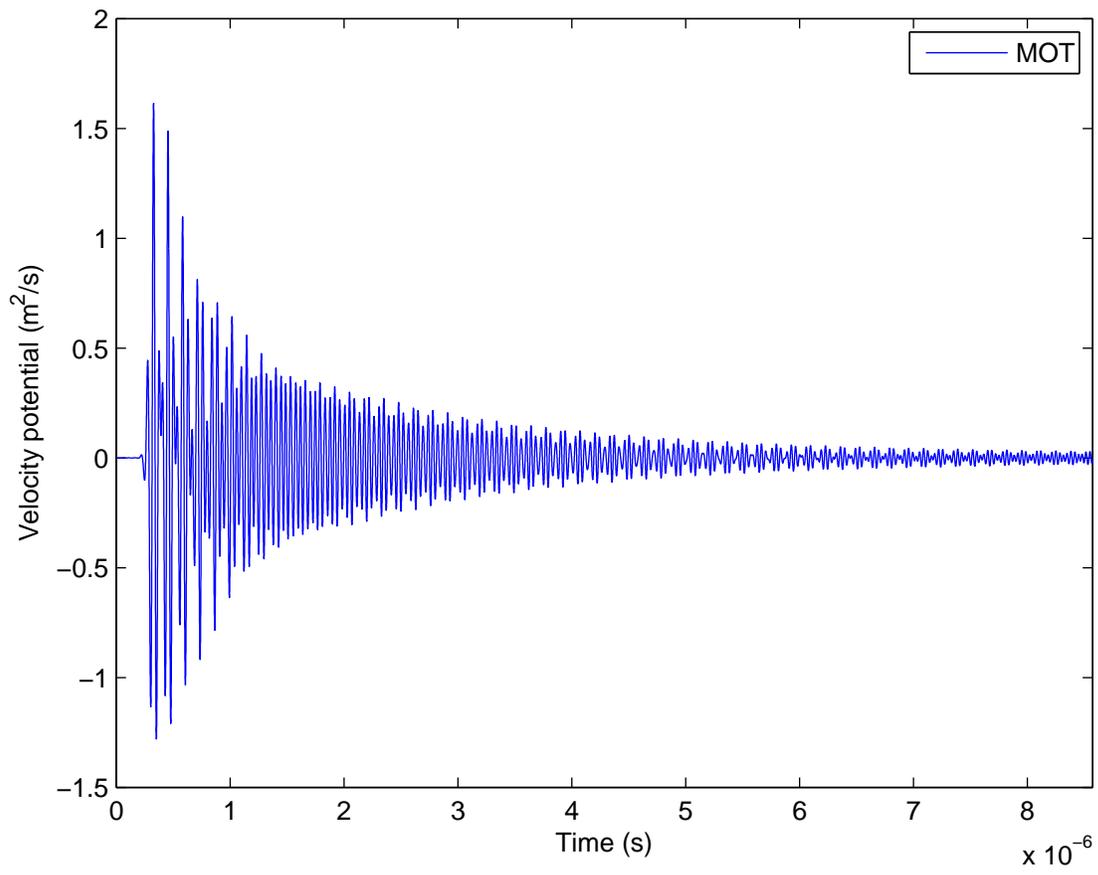
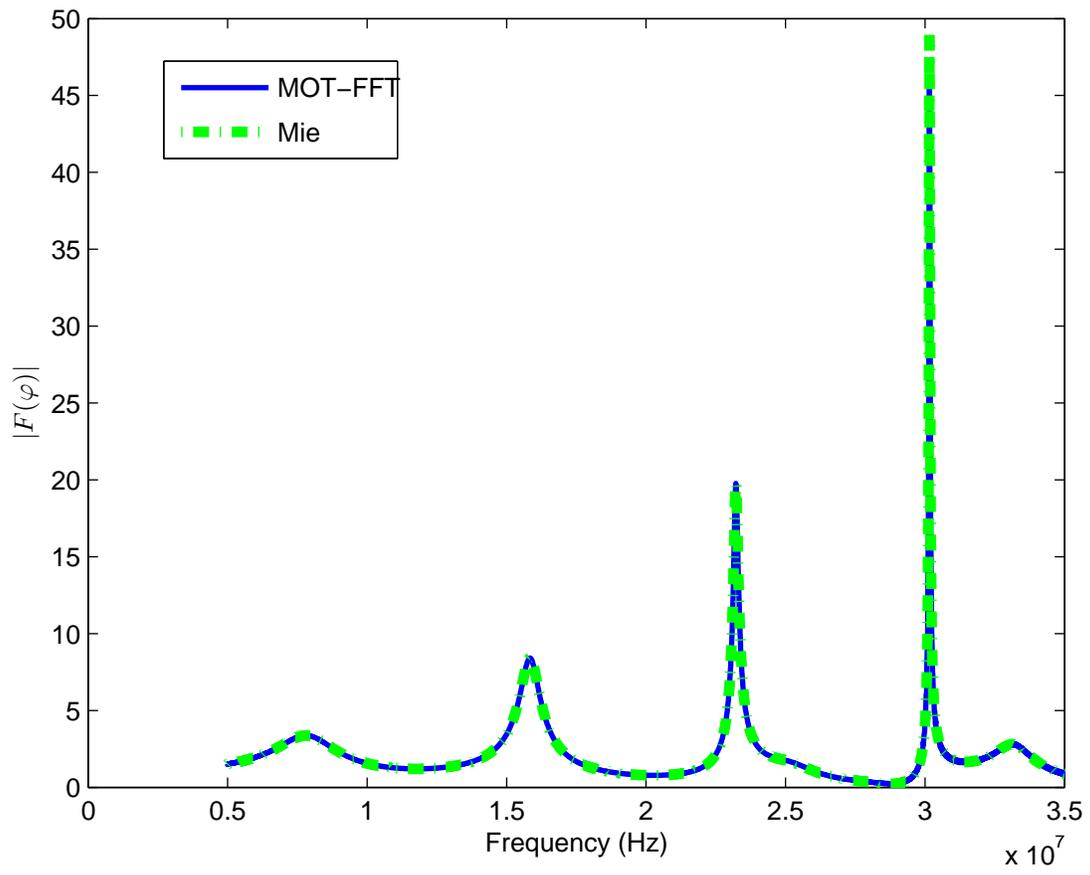
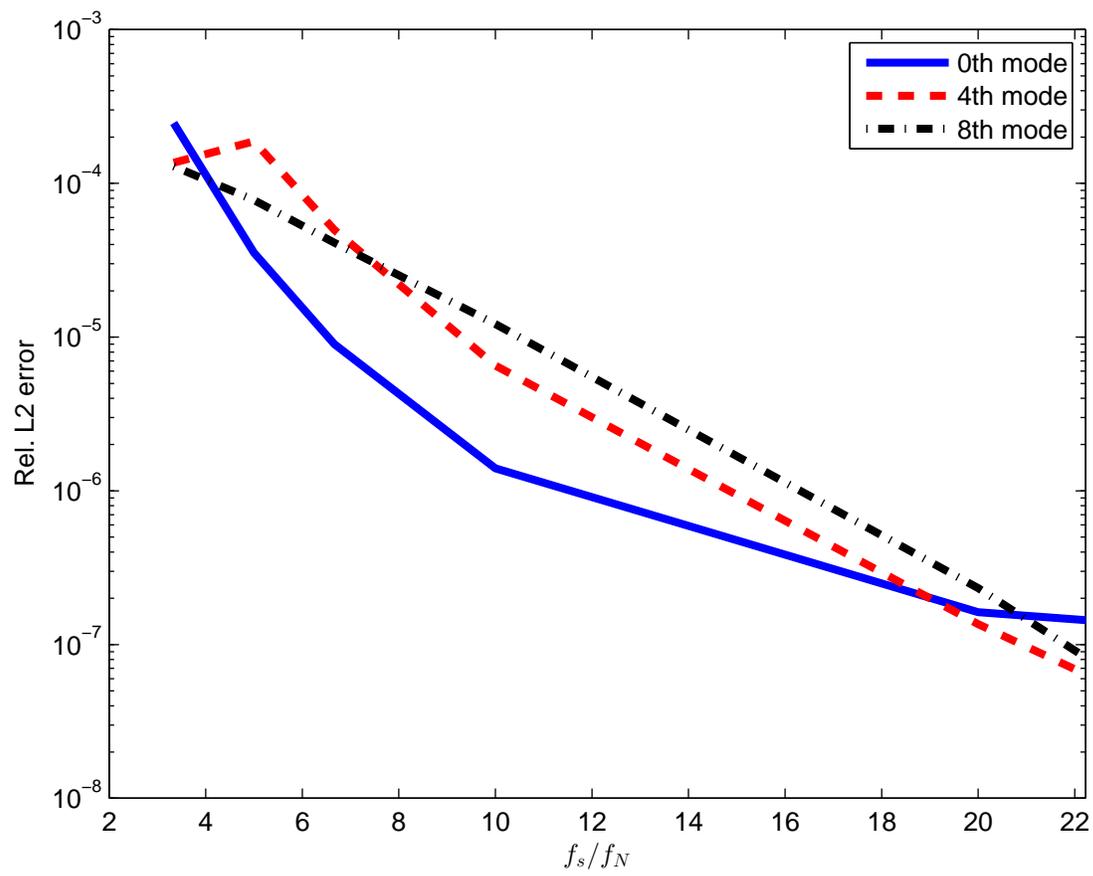
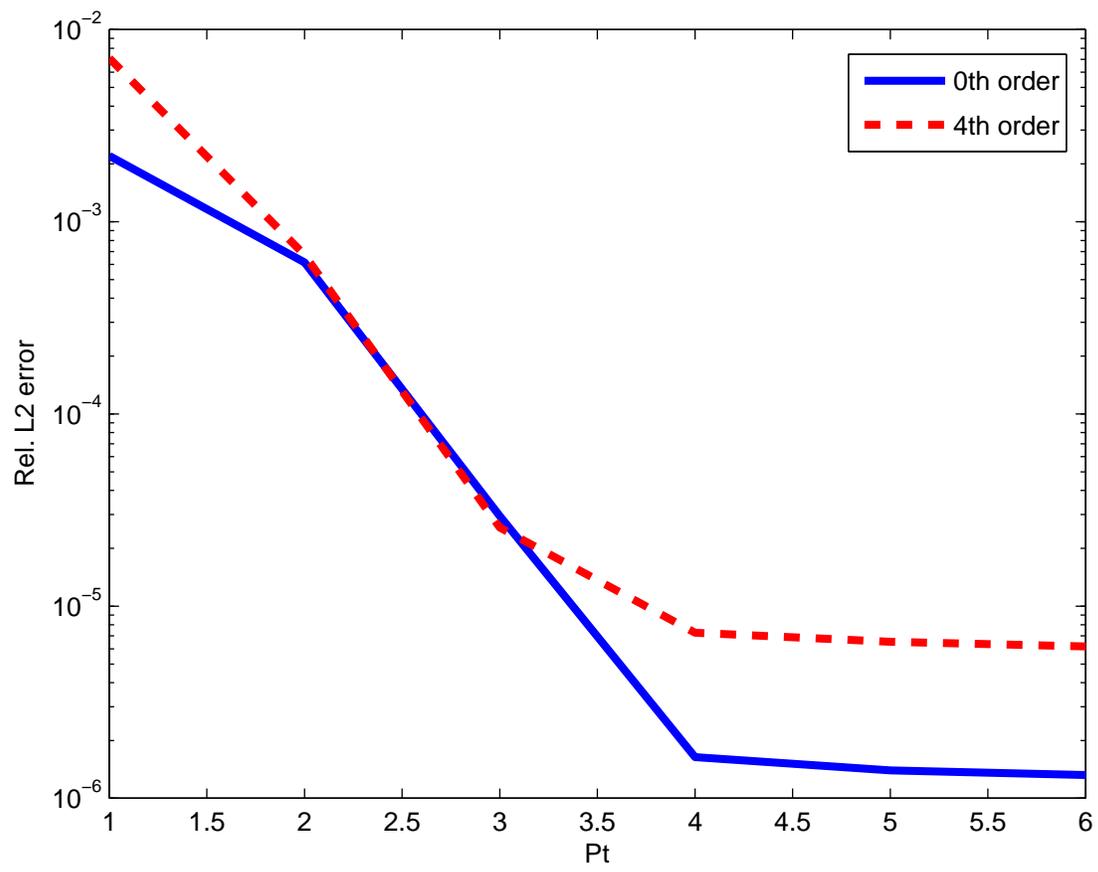
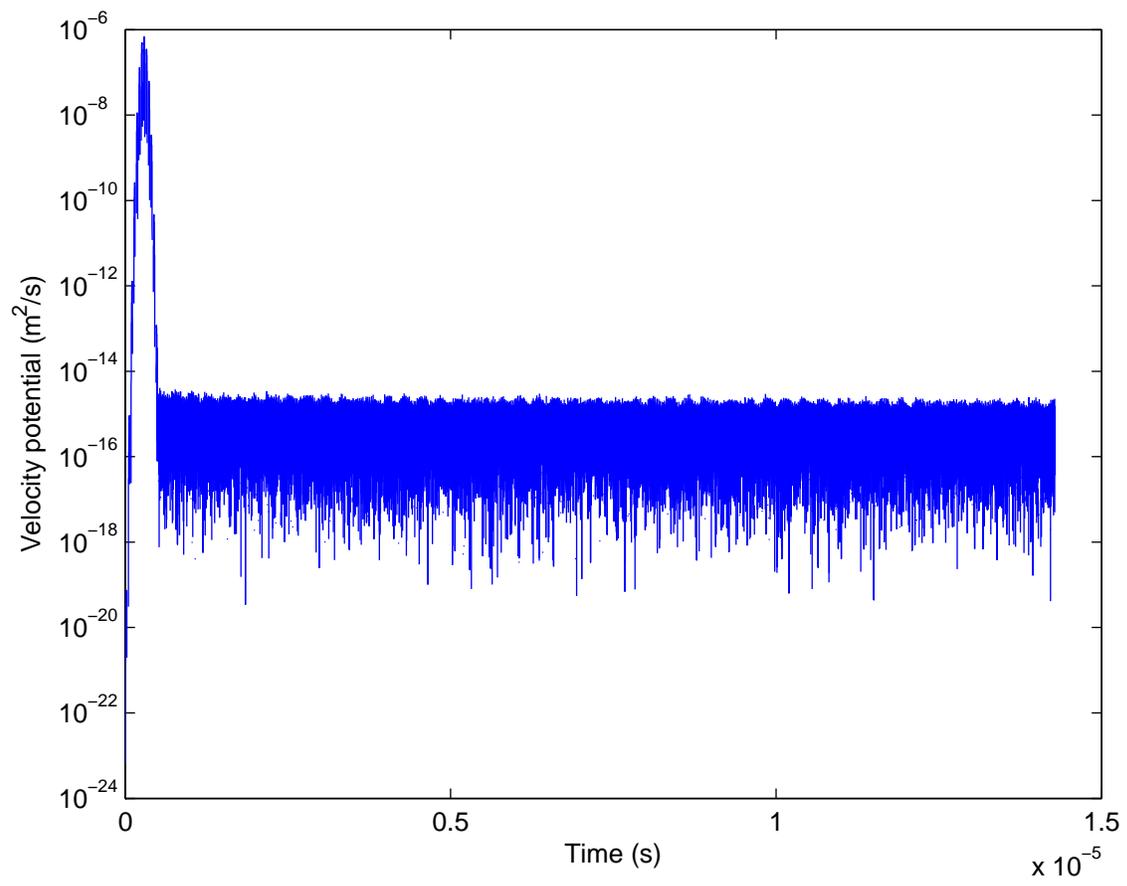
\end{document}